\documentclass[aps,prd,superscriptaddress,preprint,showpacs,amsmath,amssymb]{revtex4}
\usepackage{graphicx, bm}
\usepackage[usenames]{color}

\begin{document}

\draft
\title{Sensitivity measuring expected on the electromagnetic anomalous couplings in the $t\bar t\gamma$ vertex at the FCC-he}

\author{M. K\"{o}ksal\footnote{mkoksal@cumhuriyet.edu.tr}}
\affiliation{\small Deparment of Optical Engineering, Sivas Cumhuriyet University, 58140, Sivas, Turkey.\\}

\author{A. A. Billur\footnote{abillur@cumhuriyet.edu.tr}}
\affiliation{\small Deparment of Physics, Sivas Cumhuriyet University, 58140, Sivas, Turkey.\\}

\author{ A. Guti\'errez-Rodr\'{\i}guez\footnote{alexgu@fisica.uaz.edu.mx}}
\affiliation{\small Facultad de F\'{\i}sica, Universidad Aut\'onoma de Zacatecas\\
         Apartado Postal C-580, 98060 Zacatecas, M\'exico.\\}

\author{ M. A. Hern\'andez-Ru\'{\i}z\footnote{mahernan@uaz.edu.mx}}
\affiliation{\small Unidad Acad\'emica de Ciencias Qu\'{\i}micas, Universidad Aut\'onoma de Zacatecas\\
         Apartado Postal C-585, 98060 Zacatecas, M\'exico.\\}

\date{\today}

\begin{abstract}

In this paper, we consider the electroweak production cross-section of a single anti-top-quark, a neutrino and a photon via charged
current through the $e^-p \to e^-\bar b \to \bar t \nu_e \gamma \to \bar t(\to W^- \to (qq',  l^- \bar\nu_l)+b) \nu_e\gamma$
signal. Further, we derived the sensitivity expected to the magnetic dipole moment $(\hat a_V)$ and the electric dipole moment $(\hat a_A)$
of the top-quark at the Future Circular Collider Hadron Electron (FCC-he). We present our study for $\sqrt{s}=7.07, 10\hspace{0.8mm}TeV$,
${\cal L}=50, 100, 300, 500, 1000\hspace{0.8mm}fb^{-1}$, $\delta_{sys}=0, 3, 5\hspace{0.8mm}\%$ and $P_{e^-}=0\%, 80\%, -80\%$, respectively.
We find that the sensitivity estimated on dipole moments of the top-quark is of the order of magnitude ${\cal O}(10^{-1})$
for both hadronic and leptonic decay modes of $W^-$: $\hat a_V=[-0.2308, 0.2204]$, $|\hat a_A|=0.2259$ at $95\%$ C.L. in the hadronic
channel with unpolarized electron beam $P_{e^-}=0\%$. Our results with polarized electron beam for $P_{e^-}=80\%$ and $P_{e^-}=-80\%$
are $\hat a_V=[-0.3428, 0.3321]$, $|\hat a_A|=0.3371$ and $\hat a_V=[-0.2041, 0.1858]$, $|\hat a_A|=0.1939$ at $95\%$ C.L. in the hadronic
channel. The corresponding results for the leptonic channel with $P_{e^-}=0\%, 80\% -80\%$ are $\hat a_V=[-0.3067, 0.2963]$, $|\hat a_A|=0.3019$,
$\hat a_V=[-0.4563, 0.4456]$, $|\hat a_A|=0.4505$ and $\hat a_V=[-0.2695, 0.2512]$, $|\hat a_A|=0.2592$, respectively.
The results for $\hat a_V$ and $\hat a_A$ in the leptonic channel are weaker by a factor of 0.75 than those corresponding to the
hadronic channel. Given these prospective sensitivities we highlight that the FCC-he is potential top-quark factory that is particularly
well suited to sensitivity study on its dipole moments and with cleaner environments.

\end{abstract}


\vspace{5mm}

\maketitle

\section{Introduction}

The characteristics of the top-quark, mainly the total decay width which is one of the fundamental property of top physics
is measured with very good precision from the partial decay width $\Gamma(t \to Wb)$. In addition, its huge mass
$m_t = 173.0\pm 0.4\hspace{0.8mm}GeV$ \cite{Data2018}, as well as its anomalous couplings to bosons in the $t\bar t \gamma$,
$t\bar t Z$, $t\bar t g$, $Wtb$, $\bar t t H$ and $tq\gamma$ vertices, have turned the top-quark into one of the most attractive particles
for new physics searches. Measurements of the properties of the top-quark offer an interesting probe to understanding the electroweak sector \cite{CPYuan}
and physics Beyond the Standard Model (BSM). These and other characteristics have led to developing your own physics program for the top-quark
for present and future $pp$, $e^-p$ and  $e^+e^-$ colliders. Therefore, top-quark physics is one of the most attractive topics at the Large
Hadron Collider (LHC), as well at the High-Luminosity Large Hadron Collider (HL-LHC) and High-Energy Large Hadron Collider (HE-LHC). However,
at the post LHC era a very attractive and interesting option to study the physics of the top-quark, mainly its anomalous couplings is through
the future electron-proton $(e^-p)$ hybrid colliders, such as the Future Circular Collider Hadron Electron (FCC-he) \cite{FCChe,Fernandez,Fernandez1,Fernandez2,Huan,Acar}.
The $e^-p$ colliders will open up new perspective in the field of fundamental physics, especially for particle physics. Several potential
features in favor of this type of electron-proton colliders are the following: 1) Would represent the high resolution collider, with the
cleaner environment for exploring the substructure and dynamics inside matter, with unmatchable precision and sensitivity. 2) The center-of-mass
energies are much higher than that of the future International Linear Collider (ILC) and the Compact Linear Collider (CLIC). 3) With concurrent
$e^-p$ and $pp$ operation, the FCC-he would transform the LHC into an energy frontier accelerator facility. 4) With very precise strong
and electroweak interaction measurements and with suppressed backgrounds from strong interactions, the $e^-p$ results would make the FCC-he
a much more powerful search and measurement laboratory than present laboratories based on $pp$ collisions. 5) The joint $pp + e^-p$ facility
can become a Higgs bosons and top-quark factory for study the physics of both with an unprecedented impact. 6) For it's high-energy,
high-luminosity and while maintaining a very clean experimental environment, the FCC-he has an outstanding opportunity to discover
new physics BSM, such as Higgs sector, top-quark physics, exotic Higgs, dark matter, heavy neutrino, matter-antimatter asymmetry
and possible discoveries in QCD. All these topics are being studied very actively. In conclusion, the physics program of high-energy
and high-luminosity $e^-p$ + $pp$ collisions are vast and contain many unique opportunities. It provides high precision of Higgs boson,
top-quark, QCD and electroweak physics complementary to $e^+e^-$ colliders. Furthermore, $e^-p$ is an attractive, realistic option for
a future energy frontier collider for particle physics. For an exhaustive study on the physics and detector design concepts see Refs. \cite{FCChe,Fernandez,Fernandez1,Fernandez2,Huan,Acar}.

In the first instants of the creation of the universe, the Big Bang should have created equal amounts of matter and antimatter.
However, there is an unexplained dominance of matter over antimatter observed in the universe. CP violation offers one explanation
for the asymmetry in baryonic matter, however, there are not enough current observed sources of CP violation to account for the
total matter-antimatter asymmetry. For this reason, it is necessary to study new sources of CP violation. About this topic, the
Electric Dipole Moments (EDM) are very sensitive to CP violation in the quark and lepton sectors. EDM searches are then in the
ideal situation of that an observation in the next generation of experiments would be a clear indication of new physics BSM.
Under this perspective, the Anomalous Magnetic Dipole Moment (AMDM) and EDM of the top-quark are currently under intense scrutiny
from the point of view: theoretical, phenomenological and experimental. Within the scope of this project, CP violation can be
parameterized by the presence of anomalous couplings in the $t\bar t\gamma$ vertex of top quark production.

In this study, we focus on AMDM $(\hat a_V)$ and the EDM $(\hat a_A)$ of the top-quark. Since the AMDM and EDM of the top-quark is
chirality changing it can be significantly enhanced compared to AMDM and EDM of light fermions by the large top coupling.

For the dominant $e^-p \to e^-\bar b \to \bar t \nu_e \gamma \to \bar t(\to W^- \to (qq', l^- \bar\nu_l)+b) \nu_e\gamma$ production channel
considered here, we find that the proposed FCC-he with $\sqrt{s}=7.07, 10\hspace{0.8mm}TeV$, ${\cal L}=50, 100, 300, 500, 1000\hspace{0.8mm}fb^{-1}$
can probe the dipole moments of the top-quark with good sensitivity. We focus on two different signals: (i) the hadronic channel with
polarized electron beam for $P_{e^-}=-80, 0, 80\hspace{0.8mm}\%$, and (ii) the leptonic channel with polarized electron beam for
$P_{e^-}=-80, 0, 80\hspace{0.8mm}\%$ in the $\bar t \nu_e \gamma$ final state. We show that sensitivity measuring expected on the
electromagnetic anomalous couplings in the $t\bar t\gamma$ vertex can be probed at $95\%$ Confidence Level (C.L.) with center-of-mass
energies $\sqrt{s}=7.07, 10\hspace{0.8mm}TeV$ and integrated luminosity ${\cal L}=1000\hspace{0.8mm}fb^{-1}$ at FCC-he.

\begin{table}[!ht]
\caption{Sensitivities achievable on the electromagnetic dipole moments of the top-quark in the SM and in different processes and colliders.}
\begin{center}
\begin{tabular}{|c|c|c|c|}
\hline\hline
{\bf Model}  &    {\bf Sensitivity of the SM}                 & {\bf C. L.}  &  {\bf References}\\
\hline
SM                                                            & $a_t= 0.02$, $d_t < 10^{-30} ({\rm ecm})$   & $68 \%$  & \cite{Benreuther},
\cite{Hoogeveen,Pospelov,Soni}  \\
\hline
\hline
\hline
{\bf Model}  &    {\bf Theoretical sensitivity: $\hat a_V$, $\hat a_A$}        & {\bf C. L.}  &  {\bf Reference}\\
\hline
Top-quark pair production at LHC                              &   $ (-0.041, 0.043), (-0.035, 0.038) $ & $68 \%$    & \cite{Juste} \\
\hline
$t\bar t\gamma$ production at LHC                             &   $ (-0.2, 0.2), (-0.1, 0.1) $         & $90 \%$    & \cite{Baur} \\
\hline
Radiative $b\to s\gamma$ transitions at Tevatron and LHC      &   $ (-2, 0.3), (-0.5, 1.5) $           & $90 \%$    & \cite{Bouzas} \\
\hline
Process $pp \to p\gamma^*\gamma^*p\to pt\bar t p $ at LHC     &   $ (-0.6389, 0.0233), (-0.1158, 0.1158) $  & $68 \%$   & \cite{Sh} \\
\hline
Measurements of $\gamma p \to t\bar t$ at LHeC                &   $ (-0.05, 0.05), (-0.20, 0.20)    $       & $90 \%$   & \cite{Bouzas1} \\
\hline
Top-quark pair production $e^+e^- \to t\bar t$  at ILC        &   $ (-0.002, 0.002), (-0.001, 0.001) $       & $68 \%$  & \cite{Aguilar} \\
\hline\hline
\end{tabular}
\end{center}
\end{table}

AMDM and EDM searches of the top-quark are performed in the Standard Model (SM) and on a variety of physics processes.
The sensitivity estimated on the AMDM and EDM of the top-quark in the SM, as well as in different processes and colliders
are reported in Table I. Other direct collider probes of the AMDM and EDM have been studied extensively   \cite{Ibrahim,Atwood,Polouse,Choi,Polouse1,Aguilar0,Amjad,Juste,Asner,Abe,Aarons,Brau,Baer,Grzadkowski:2005ye,murat,Billur}.

Plan of the article is as follows: In Section II, we introduce the top-quark effective electromagnetic interactions. In Section III,
we sensitivity measurement on top-quark anomalous electromagnetic couplings through $e^-p \to e^-\bar b \to \bar t \nu_e \gamma \to
\bar t(\to W^- \to (qq', l^- \bar\nu_l)+b) \nu_e\gamma$ signal. Finally, we present our conclusions in Section IV.

\section{Single top-quark production via the process $e^-p \to e^-\bar b \to \bar t \nu_e \gamma$}

\subsection{Effective Lagrangian of $t\bar t \gamma$ interaction  of the top-quark}

The SM predicts CP violation outside the K, D and B meson systems is small to be observed. However, in some extensions of the SM,
CP violation might be considerably enhanced, especially in the presence of heavy particles as the top quark. In particular,
CP-violating EDM of the top-quark could be enhanced.

Single top-quark production processes are sensitive to the anomalous couplings in the $t\bar t \gamma$ vertex. Furthermore,
since the top-quark lifetime is shorter than the timescale of spin decoherence induced by QCD, its decay products preserve
information of its polarization by the production mechanism. This provides additional powerful tools in the search
for BSM physics in single top-quark studies.

On the other hand, due to the absence so far of any signal of new heavy particles decaying into top-quark, an attractive
approach for describing possible new physics effects in a model-independent way is based on effective Lagrangian. In this
approach, all the heavy degrees of freedom are integrated out leading to obtain the effective interactions between the SM
particles. This is justified because the related observables have not shown any significant deviation from the SM predictions
so far. The Lagrangian describing interaction of the anomalous $tt\gamma$ coupling including the SM contribution and terms BSM
which are related to new physics has the structure:

\begin{equation}
{\cal L}_{eff}={\cal L}^{(4)}_{SM} + \frac{1}{\Lambda^2}\sum_n \Bigl[C_n{\cal O}^{(6)}_n + C^*_n{\cal O}^{\dagger(6)}_n \Bigr].
\end{equation}

\noindent Here, ${\cal L}_{eff}$ is the effective Lagrangian gauge-invariant which contains a series of dimension-six operators
built with the SM fields, ${\cal L}^{(4)}_{SM}$ is the renormalizable SM Lagrangian of dimension-four, $\Lambda$ is the scale at
which new physics expected to be observed, $C_n$ are Wilson coefficients which are dimensionless coefficients and ${\cal O}^{(6)}_n$
represents the dimension-six gauge-invariant operator. The ${\cal O}^{(6)}_n$ operator and the unknown coefficients $C_n$, combined
with $\Lambda^{-2}$, produce the non-standard coupling constants, that is generated anomalous contributions to the photon-top-quark
interaction vertex which is similar in structure to radiative corrections in the SM.

The most general Lagrangian term that one can write for the $t\bar t\gamma$ coupling up to dimension-six gauge invariant
operators \cite{Sh,Kamenik,Baur,Aguilar,Aguilar1} is:

\begin{equation}
{\cal L}_{t\bar t\gamma}=-g_eQ_t\bar t \Gamma^\mu_{ t\bar t  \gamma} t A_\mu,
\end{equation}

\noindent this equation includes the SM coupling and contributions from dimension-six effective operators. $g_e$ is the
electromagnetic coupling constant, $Q_t$ is the top-quark electric charge and the Lorentz-invariant vertex function
$\Gamma^\mu_{t\bar t \gamma}$ is given by:

\begin{equation}
\Gamma^\mu_{t\bar t\gamma}= \gamma^\mu + \frac{i}{2m_t}(\hat a_V + i\hat a_A\gamma_5)\sigma^{\mu\nu}q_\nu.
\end{equation}

\noindent $m_t$ and $q$ are the mass of the top-quark and the momentum transfer to the photon, respectively. The $\hat a_V$
and $\hat a_A$ couplings in Eq. (3) are real and related to the AMDM and the EDM of the top-quark. These couplings
$\hat a_V$ and $\hat a_A$ are directly related to the $a_t$ and $d_t$, via the relations:

\begin{eqnarray}
\hat a_V&=&Q_t a_t,  \\
\hat a_A&=&\frac{2m_t}{e}d_t.
\end{eqnarray}

As shown in Refs. \cite{Buhmuller,Aguilar2,Antonio,Grzadkowski:2005ye}, among operators of dimension-six there exist only two relevant
for the $t\bar t\gamma$ interaction operator:

\begin{eqnarray}
{\cal O}_{uW}^{33}&=&\bar q_{L3}\sigma^{\mu\nu}\tau^a t_{R}{\tilde \phi} W_{\mu\nu}^{a}+{\mbox{h.c.}},\\
{\cal O}_{uB\phi}^{33}&=&\bar q_{L3}\sigma^{\mu\nu}t_{R}{\tilde \phi} B_{\mu\nu}+{\mbox{h.c.}},
\end{eqnarray}

\noindent where the index 3 means the 3rd quark generation, $\bar q_{L3}$ is the quark field, $\sigma^{\mu\nu}$ are the
Pauli matrices, ${\tilde \phi}=i\tau_2\phi^*$, $\phi$ is the SM Higgs doublet, $W_{\mu\nu}^{a}$ and $B_{\mu\nu}$
are the $U(1)_Y$ and $SU(2)_L$ gauge field strength tensors which are defined as:

\begin{eqnarray}
B_{\mu\nu}&=&\partial_\mu B_\nu-\partial_\nu B_\mu,  \\
W^a_{\mu\nu}&=&\partial_\mu W^a_\nu-\partial_\nu W^a_\mu-g\epsilon^{abc}W^b_\mu W^c_\nu,
\end{eqnarray}

\noindent with $a,b,c=1,2,3$. From the parametrization given by Eq. (3), and from the operators of dimension-six given in
Eqs. (6) and (7) after replacing $<{\tilde \phi}> \to \frac{1}{\sqrt{s}}$ give rise to the corresponding CP even ${\hat a_V}$
and CP odd ${\hat a_A}$ couplings:

\begin{eqnarray}
\hat a_V=\frac{2 m_t}{e} \frac{\sqrt{2}\upsilon}{\Lambda^{2}} Re\Bigl[\cos\theta _{W} C_{uB\phi}^{33} + \sin\theta _{W} C_{uW}^{33}\Bigr],\\
\hat a_A=\frac{2 m_t}{e} \frac{\sqrt{2}\upsilon}{\Lambda^{2}} Im\Bigl[\cos\theta _{W} C_{uB\phi}^{33} + \sin\theta _{W} C_{uW}^{33}\Bigr],
\end{eqnarray}

\noindent which are related to the AMDM and EDM of the top-quark. The ${\hat a_V}$ and ${\hat a_A}$ couplings contain $\upsilon=246$ GeV,
the breaking scale of the electroweak symmetry, $\sin\theta _{W} (\cos\theta _{W})$, the sine(cosine) of the weak mixing angle
and $\Lambda$ is the new physics scale.

\subsection{Cross-section of the $e^-p \to e^-\bar b \to \bar t \nu_e \gamma \to \bar t(\to W^- \to (qq',  l^- \bar\nu_l)+b) \nu_e\gamma$ signal}

The FCC-eh project \cite{FCChe,Fernandez,Fernandez1,Fernandez2,Huan,Acar} offer $e^-p$ collisions at TeV-scale center-of-mass energy
and luminosities of order 1000 times larger than that of Hadron-Electron Ring Accelerator (HERA), the first and to date the only
lepton-hadron collider worldwide, providing fascinating probes of QCD and hadron structure as well as a novel configuration for Higgs
boson, top-quark and BSM physics.

The physics highlights are available from combining the capabilities of these facilities they were already mentioned in the introduction.
However, a general aspect is to maximize the BSM physics search potential at high energies by exploiting the unique capabilities of
an $e^-p$ collider.

The most significant top-quark production processes at the $e^-p$ colliders are single top-quark, $t\bar t$, and associated $tW$
production. In Ref. \cite{Bouzas1} is shown the values of the associated cross-sections for these processes. The main source of
production is single-top via the charged current $W$ t-channel \cite{Moretti}, whereas for the signals, $t\bar t$ and $tW$, the
cross-section is minor. In addition, taking into account the advantage of an experimental cleaner environment than the $pp$ colliders,
we can anticipate a potential efficiency of these colliders to study the top-quark physics.

The deep inelastic $e^-p \to e^-\bar b \to \bar t \nu_e \gamma \to \bar t(\to W^- \to (qq',  l^- \bar\nu_l)+b) \nu_e\gamma$
scattering process is measured at FCC-he via the exchange of a $W^{\pm}$ boson in charged current scattering
as shown in Figs. 1 and 2.

For the calculation of the charged current cross-section, we consider the CTEQ6L1 PDFs \cite{CTEQ6L1} and we apply the following detector
acceptance cuts on the pseudorapidity of the photon and the transverse momentum of the photon and the neutrino, respectively, to reduce the
background and to optimize the signal sensitivity:

\begin{eqnarray}
|\eta^{\gamma}|&<& 2.5,   \nonumber \\
p^\gamma_T &>& 20\hspace{0.8mm}GeV,  \\
p^{(\nu)}_T &>& 20\hspace{0.8mm}GeV. \nonumber
\end{eqnarray}

The proposed FCC-he is well-suited for discovering physics BSM and for precisely unraveling the structure of the fundamental physics
with unpolarized and polarized electron beam. In particle physics, polarization refers to the extent to which a particle spin is aligned
along a particular direction. The t-channel single-top-quarks of the process $e^-p \to e^-\bar b \to \bar t \nu_e \gamma$ are produced
with a strong degree of polarization along the direction of the momentum of the light spectator quark, whose direction then defines the
top-quark spin axis. Regarding this subject, the physics in study can be maximized by the use of polarized beams. In this paper shows
the important role of polarized beam and summarizes the benefits obtained from polarizing the electron beam. The polarized $e^-$ beam,
combined with the clean experimental environment provided by the FCC-he, will allow to improve strongly the potential of searches for the
dipole moments, which opens the possibility to resolve shortcomings of the SM. With these arguments, we consider polarized
electron beam in our study. The formula for the total cross-section for an arbitrary degree of longitudinal $e^-$ beams
polarization is given by \cite{XiaoJuan}:

\begin{eqnarray}
\sigma_{e^-_r}= \sigma_{e^-_0}\cdot (1-P_{e^-_r}),  \hspace{7mm}
\sigma_{e^-_l} + \sigma_{e^-_r} &=& 2\sigma_{e^-_0},
\end{eqnarray}

\noindent where $\sigma_{e^-_r}$, $\sigma_{e^-_l}$ and $\sigma_{e^-_0}$ represent the right, left and without electron
beam polarization, respectively and $P_{e^-}$ is the polarization degree of the electron.

We have implemented $t\bar t \gamma$ effective coupling corresponding given by the Lagrangian (2) in CalcHEP \cite{Belyaev}
to compute the tree level amplitudes relevant for the process. The partonic cross-section is convoluted with CTEQ6L1
PDFs \cite{CTEQ6L1}. Finally, we use CalcHEP to compute numerically the cross-section $\sigma(\sqrt{s}, \hat a_V, \hat a_A, P_{e^-})$
as a function of the center-of-mass energy and effective couplings. We displayed the $7.07\hspace{0.8mm}TeV$ and $10\hspace{0.8mm}TeV$
cross-section of the $2 \to 3$ process $e^-p \to e^-\bar b \to \bar t \nu_e \gamma$ with $P_{e^-}=0\%$, $P_{e^-}=-80\%$ and
$P_{e^-}=80\%$ in Eqs. (14)-(25) for the FCC-he:\\

$i)$ Total cross-section for $\sqrt{s}=7.07\hspace{0.8mm} TeV$ and $P_{e^-}=0\%$:

\begin{eqnarray}
\sigma(\hat a_V)&=&\Bigl[(0.0236)\hat a^2_V +(0.0000489)\hat a_V  +0.737 \Bigr] (pb),   \\
\sigma(\hat a_A)&=&\Bigl[(0.0236)\hat a^2_A + 0.737 \Bigr] (pb).
\end{eqnarray}

$ii)$ Total cross-section for $\sqrt{s}=10\hspace{0.8mm} TeV$ and $P_{e^-}=0\%$:

\begin{eqnarray}
\sigma(\hat a_V)&=&\Bigl[(0.0593)\hat a^2_V +(0.000618)\hat a_V  + 1.287 \Bigr] (pb),   \\
\sigma(\hat a_A)&=&\Bigl[(0.0593)\hat a^2_A + 1.287 \Bigr] (pb).
\end{eqnarray}

$iii)$ Total cross-section for $\sqrt{s}=7.07\hspace{0.8mm} TeV$ and $P_{e^-}=80\%$:

\begin{eqnarray}
\sigma(\hat a_V)&=&\Bigl[(0.00472)\hat a^2_V +(0.0000109)\hat a_V  +0.148 \Bigr] (pb),   \\
\sigma(\hat a_A)&=&\Bigl[(0.00472)\hat a^2_A + 0.148 \Bigr] (pb).
\end{eqnarray}

$iv)$ Total cross-section for $\sqrt{s}=10\hspace{0.8mm} TeV$ and $P_{e^-}=80\%$:

\begin{eqnarray}
\sigma(\hat a_V)&=&\Bigl[(0.0119)\hat a^2_V +(0.000127)\hat a_V  +0.257 \Bigr] (pb),   \\
\sigma(\hat a_A)&=&\Bigl[(0.0119)\hat a^2_A + 0.257 \Bigr] (pb).
\end{eqnarray}

$v)$ Total cross-section for $\sqrt{s}=7.07\hspace{0.8mm} TeV$ and $P_{e^-}=-80\%$:

\begin{eqnarray}
\sigma(\hat a_V)&=&\Bigl[(0.0423)\hat a^2_V +(0.000417)\hat a_V  + 1.328 \Bigr] (pb),   \\
\sigma(\hat a_A)&=&\Bigl[(0.0423)\hat a^2_A + 1.328 \Bigr] (pb).
\end{eqnarray}

$vi)$ Total cross-section for $\sqrt{s}=10\hspace{0.8mm} TeV$ and $P_{e^-}=-80\%$:

\begin{eqnarray}
\sigma(\hat a_V)&=&\Bigl[(0.107)\hat a^2_V +(0.00196)\hat a_V  +2.315 \Bigr] (pb),   \\
\sigma(\hat a_A)&=&\Bigl[(0.107)\hat a^2_A + 2.315 \Bigr] (pb).
\end{eqnarray}

Our results given by Eqs. (14)-(25) show the effect of taking $-80\%$ beam polarization for electron, which results in the
enhancement of the SM and non-SM single-top production cross-section as the cross-section scales as $(1 + P_{e^-})$, $P_{e^-}$
being the degree of polarization of the electron.

The variation of the single top-quark production cross-section with the effective $t\bar t \gamma$ couplings, $\hat a_V$
or $\hat a_A$ and taking one anomalous coupling at a time are shown in Figs. 3-6. The curves depict the cross-section for
$e^-p \to e^-\bar b \to \bar t \nu_e \gamma$ from the $80\%$, $-80\%$ polarized and unpolarized $e^-$ beam, respectively,
and we fixed the energy of the $e^-$ beam $E_e = 250 \hspace{0.8mm}GeV$, $500\hspace{0.8mm}GeV$ and the energy of the $p$
beam $E_p = 50\hspace{0.8mm}TeV$. These figures have shown
a stronger dependence of the cross-section on the anomalous coupling $\hat a_V(\hat a_A)$ in the range allowed by these
parameters. Our results indicate that considering the proposed $10\hspace{0.8mm}TeV$ energy, detector acceptance cuts (see Eq. (12))
and $-80\%$ electron polarization, the cross-sections as a function of $\hat a_V$ or $\hat a_A$ are higher.
For instance, the cross-section projected is $\sigma(\hat a_V, -80\%)=(1.66)\sigma(\hat a_V, 0\%)$ for $\sqrt{s}= 7.07\hspace{0.8mm}TeV$,
while $\sigma(\hat a_V, -80\%)=(1.78)\sigma(\hat a_V, 0\%)$ for $\sqrt{s}= 10\hspace{0.8mm}TeV$, that is, there is an improvement
in the cross-section by a factor of 1.66 (1.78) for the polarized case with respect to the case unpolarized. Similar results are
obtained for $\sigma(\hat a_A, 80\%)$.

The cross-sections for the energies $\sqrt{s}= 7.07\hspace{0.8mm}TeV$ and $10\hspace{0.8mm}TeV$ are shown in Figs. 7 and 8.
As can be seen from Figs. 7-8, the surfaces $\sigma(e^-p \to \bar t \nu_e \gamma)$ as functions of $\hat a_V$ and $\hat a_A$
have extreme points: maximum and minimum. The minimum value corresponds to the SM, while the maximum value corresponds to the
anomalous contribution, which is consistent with Eqs. (14)-(25). In both figures, the cross-section depends significantly on
the observables $\hat a_V$ and $\hat a_A$.

With the purpose of comparison and analysis, we compare our results for the anomalous couplings $\hat a_V$ and $\hat a_A$ with those
quoted in the papers \cite{Sh,Bouzas} (see also Table I). The authors of Ref. \cite{Bouzas}, specifically discussed the bounds on
the AMDM and EDM of the top-quark that can be obtained from measurements of the semi-inclusive decays $B \to X_s\gamma$, and of
$t\bar t\gamma$ production at the Tevatron and the LHC. Performing their analysis they find that the AMDM is bounded by
$-2 < \kappa < 0.3$  whereas the EDM is bound by $-0.5 < \tilde\kappa < 1.5$, respectively. For our case, we consider the process
of single anti-top-quark production through charged current with the $e^-p \to e^-\bar b \to \bar t \nu_e \gamma \to \bar t(\to W^- \to
(qq',  l^- \bar\nu_l)+b) \nu_e\gamma$ signal. We based our results on the data at $\sqrt{s}=10\hspace{0.8mm}TeV$,
${\cal L}=1000\hspace{0.8mm}fb^{-1}$, $\delta_{sys}=0\%$, $P_{e^-}=0\%$ and $95\%\hspace{0.8mm}C.L.$, we obtain $\hat a_V=(-0.2308, 0.2204)$,
$\hat a_A=|0.2259|$ and $\hat a_V=(-0.3067, 0.2963)$, $\hat a_A=|0.3019|$ for the hadronic and leptonic modes. Although the conditions
for the study of the dipole moments of the top-quark through the $b\to s\gamma$ transitions at Tevatron and LHC, and
$e^-p \to e^-\bar b \to \bar t \nu_e \gamma \to \bar t(\to W^- \to (qq',  l^- \bar\nu_l)+b) \nu_e\gamma$
are different, our results are competitive with respect to the results reported in Ref. \cite{Bouzas}.
More recently, using the process $pp\to p\gamma^* \gamma^* p \to pt\bar t p$, a detailed study on the top-quark anomalous couplings
$\hat a_v$ and $\hat a_A$ for LHC at $14\hspace{0.8mm}TeV$ with $300\hspace{0.8mm} fb^{-1}$ of data \cite{Sh} is done. The $68\%$ C.L.
bounds that they have obtained are found to be in the intervals of $\hat a_V= (−0.6389, 0.0233)$ and $\hat a_A=(−0.1158, 0.1158)$.
From the comparison of our study via the process $e^-p \to e^-\bar b \to \bar t \nu_e \gamma \to \bar t(\to W^- \to (qq',  l^- \bar\nu_l)+b) \nu_e\gamma$
at the FCC-he, with respect to the process $pp\to p\gamma^* \gamma^* p \to pt\bar t p$ at the LHC, our results indicate a significant improvement
in the measurements for $\hat a_V$ and $\hat a_A$. Additionally, it is noteworthy that with our process the total cross-sections is a factor
${\cal O}(10^3)$ between $pp\to p\gamma^* \gamma^* p \to pt\bar t p$ and $e^-p \to e^-\bar b \to \bar t \nu_e \gamma \to \bar t(\to W^- \to (qq',  l^- \bar\nu_l)+b) \nu_e\gamma$, indicating that our results project 3 orders of magnitude more higher than the reported in Ref. \cite{Sh}.
These predictions indicate that the sensitivity on the anomalous couplings $\hat a_V$ and $\hat a_A$ can be measured better at the FCC-he
by a few orders of magnitude in comparison with the predictions of the LHC.

\section{Model-independent sensitivity estimates on the $\hat a_V$ and $\hat a_A$}

In Tables II-VII is shown the results for the  model-independent sensitivity achievable at $95\%$ C.L. for the non-standard couplings
$\hat a_V$ and $\hat a_A$ obtained from an analysis of the process $e^-p \to e^- \bar b \to \bar t \nu_e \gamma$ at the FCC-he.
At the FCC-he, we assume the center-of-mass energies $\sqrt{s}=7.07, 10\hspace{0.8mm}TeV$ and luminosities ${\cal L}= 50, 100, 300, 500, 1000\hspace{0.8mm}fb^{-1}$ with unpolarized and
polarized electron beam $P_{e^-}=-80\%, 0\%, 80\%$. Additionally, we impose the acceptance cuts for the FCC-he given by Eq. (12) and
take into account the systematic uncertainties $\delta_{sys}=0\%, 3\%, 5\%$.

In order to extract the expected sensitivity at $95\%$ C.L. on the effective operators couplings $\hat a_V$ and $\hat a_A$, we compute
$\sigma_{BSM}(\hat a_V, \hat a_A)$ cross-section of the process $e^-p \to e^- \bar b \to \bar t \nu_e \gamma$ as function of the effective
couplings as discussed in the previous section, and we assume the measured cross-section to coincide with the SM predictions and we construct
the following $\chi^2$ function:

 \begin{equation}
\chi^2(\hat a_V, \hat a_A)=\biggl(\frac{\sigma_{SM}-\sigma_{BSM}(\sqrt{s}, \hat a_V, \hat a_A, P_{e^-})}{\sigma_{SM}\sqrt{(\delta_{st})^2
+(\delta_{sys})^2}}\biggr)^2.
\end{equation}

\noindent $\sigma_{SM}$ is the cross-section of the SM and $\sigma_{BSM}(\sqrt{s}, \hat a_V, \hat a_A, P_{e^-})$ is the total cross-section
containing contributions from the SM and BSM, while $\delta_{st}=\frac{1}{\sqrt{N_{SM}}}$ and $\delta_{sys}$ are the statistical and systematic
uncertainties, respectively. The number of events $N_{SM}$ for the process $e^-p \to e^- \bar b \to \bar t \nu_e \gamma$
is calculated by $N_{SM}={\cal L}_{int}\times \sigma_{SM} \times BR(\bar t \to W^-b)\times BR(W^- \to qq' (l^-\nu_l))\times  \epsilon_{b-tag}$, where ${\cal L}_{int}$ is the integrated FCC-he luminosity and  $b$-jet tagging efficiency is $\epsilon_b=0.8$ \cite{atlas}.
The top-quark decays weakly and almost $100\%$ to a $W$ boson and $b$ quark, specifically $\bar t\to \bar bW^-$, where the $W$
boson decays to either hadronically ($W \to qq'$) or leptonically ($W\to l^-\nu_l$), with a Branching Ratio of: $BR(W \to q q')=0.674$
for hadronic decay, $BR(W \to l\nu_l)(l=e, \mu)=0.213$ for light leptonic decays and $BR(W \to \tau\nu_\tau)=0.113$ \cite{Data2018}.

Next, we present the sensibility measurement for the anomalous couplings $\hat a_V$ and $\hat a_A$ as is shown in Tables II-VII
which are obtained for $\sqrt{s}=7.07, 10\hspace{0.8mm}TeV$, ${\cal L}= 50-1000\hspace{0.8mm}fb^{-1}$ and $P_{e^-}=-80\%, 0\%, 80\%$,
where only one coupling at a time is varied.

From Tables II-VII, the results for the dipole moments $\hat a_V$ and $\hat a_A$, for specific values of $\sqrt{s}= 10\hspace{0.8mm}TeV$,
${\cal L}=1000\hspace{0.8mm}fb^{-1}$, $P_{e^-}= -80\%, 0\%, 80\%$ and $\delta_{sys}=0\%$ are as follows:\\

$i)$ Sensitivity on $\hat a_V$ and $\hat a_A$ for $\sqrt{s}=10\hspace{0.8mm} TeV$, $P_{e^-}=-80\%$ and BR($W^- \to$ hadronic):

\begin{eqnarray}
-0.2041 \leq & \hat a_V & \leq 0.1858, \hspace{3mm}   \mbox{$95\%$ C.L.}, \\
-0.1939 \leq & \hat a_A & \leq 0.1939, \hspace{3mm}   \mbox{$95\%$ C.L.}.
\end{eqnarray}


$ii)$ Sensitivity on $\hat a_V$ and $\hat a_A$ for $\sqrt{s}=10\hspace{0.8mm} TeV$, $P_{e^-}=-80\%$ and BR($W^- \to$ leptonic):

\begin{eqnarray}
-0.2695 \leq & \hat a_V & \leq 0.2512, \hspace{3mm}   \mbox{$95\%$ C.L.}, \\
-0.2592 \leq & \hat a_A & \leq 0.2592, \hspace{3mm}   \mbox{$95\%$ C.L.}.
\end{eqnarray}

$iii)$ Sensitivity on $\hat a_V$ and $\hat a_A$ for $\sqrt{s}=10\hspace{0.8mm} TeV$, $P_{e^-}=0\%$ and BR($W^- \to$ hadronic):

\begin{eqnarray}
-0.2308 \leq & \hat a_V & \leq 0.2204, \hspace{3mm}   \mbox{$95\%$ C.L.}, \\
-0.2259 \leq & \hat a_A & \leq 0.2259, \hspace{3mm}   \mbox{$95\%$ C.L.}.
\end{eqnarray}

$iv)$ Sensitivity on $\hat a_V$ and $\hat a_A$ for $\sqrt{s}=10\hspace{0.8mm} TeV$, $P_{e^-}=0\%$ and BR($W^- \to$ leptonic):

\begin{eqnarray}
-0.3067 \leq & \hat a_V & \leq 0.2963, \hspace{3mm}   \mbox{$95\%$ C.L.}, \\
-0.3019 \leq & \hat a_A & \leq 0.3019, \hspace{3mm}   \mbox{$95\%$ C.L.}.
\end{eqnarray}

$v)$ Sensitivity on $\hat a_V$ and $\hat a_A$ for $\sqrt{s}=10\hspace{0.8mm} TeV$, $P_{e^-}=80\%$ and BR($W^- \to$ hadronic):

\begin{eqnarray}
-0.3428 \leq & \hat a_V & \leq 0.3321, \hspace{3mm}   \mbox{$95\%$ C.L.}, \\
-0.3371 \leq & \hat a_A & \leq 0.3371, \hspace{3mm}   \mbox{$95\%$ C.L.}.
\end{eqnarray}

$vi)$ Sensitivity on $\hat a_V$ and $\hat a_A$ for $\sqrt{s}=10\hspace{0.8mm} TeV$, $P_{e^-}=80\%$ and BR($W^- \to$ leptonic):

\begin{eqnarray}
-0.4563 \leq & \hat a_V & \leq 0.4456, \hspace{3mm}   \mbox{$95\%$ C.L.}, \\
-0.4505 \leq & \hat a_A & \leq 0.4505, \hspace{3mm}   \mbox{$95\%$ C.L.}.
\end{eqnarray}

A direct comparison of Eqs. (27)-(38) for $\hat a_V$ and $\hat a_V$ shown that the sensitivity is increases up to $12\%$ for the case
with $P_{e^-}=- 80\%$ and BR($W^- \to$ hadronic, leptonic) that for the case with $P_{e^-}=0\%$ and BR($W^- \to$ hadronic, leptonic).
While from Eqs. (27)-(30) and (35)-(38) the sensitivity is increases up to $67\%$ for the case with $P_{e^-}=-80\%$ and
BR($W^- \to$ hadronic, leptonic) with respect to the case with $P_{e^-}=80\%$ and BR($W^- \to$ hadronic, leptonic), respectively.

For $P_{e^-}=0\%$ and the hadronic channel of the W boson, Figs. 9-10 show the prospects of the sensitivity on the electromagnetic anomalous
couplings $\hat a_V$ and $\hat a_V$ at the FCC-he. Also, in order to obtain the plots we have assumed that $\sqrt{s}=7.07, 10\hspace{0.8mm}TeV$
and ${\cal L}= 50, 250, 1000\hspace{0.8mm}fb^{-1}$ at the $95\%$ C.L.. Of these contours plots, our forecast for the future sensitivity of the
observables $\hat a_V$ and $\hat a_A$ are based on the process $e^-p \to e^- \bar b \to \bar t \nu_e \gamma$, as well as in the future projections
of the FCC-he for $\sqrt{s}$ and ${\cal L}$. The regions allowed for the top-quark AMDM, EDM and collider sensitivity are colored in
pink, blue, purple, respectively, while the prediction corresponds to the SM can be obtained from Eqs. (14)-(25). The $95\%$ C.L. regions
for each of these couplings separately are $\hat a_V  \hspace{1mm}\in \hspace{1mm}[-0.25, 0.25]$, $\hat a_A \hspace{1mm}\in \hspace{1mm}[-0.29, 0.29]$
for $\sqrt{s}=7.07\hspace{0.8mm}TeV$ and ${\cal L}= 250 \hspace{0.8mm}fb^{-1}$. In addition, $\hat a_V  \hspace{1mm}\in \hspace{1mm}[-0.20, 0.20]$,
$\hat a_A \hspace{1mm}\in \hspace{1mm}[-0.20, 0.20]$ for $\sqrt{s}=10\hspace{0.8mm}TeV$ and  ${\cal L}= 1000 \hspace{0.8mm}fb^{-1}$. These results
are consistent with those shown in Tables IV and V. It is worth mentioning that, the results obtained in Tables IV and V, as well as the corresponding
results obtained through the contours Figs. 9 and 10, in some cases are more sensitive than the reported ones in Table I. In particular, an improvement
is reachable in comparison with the constraints obtained from the radiative $b \to s\gamma$ transitions at Tevatron and LHC \cite{Bouzas} and
$pp\to p\gamma^* \gamma^* p \to pt\bar t p$ \cite{Sh} searches mentioned in Table I and subsection B.

\section{Conclusions}

As mentioned above, due to its sizable $t\bar t\gamma$ coupling the top-quark is one of the most attractive particles and provides
one of the most convincing  alternatives to probe new physics BSM, such as AMDM ($\hat a_V$) and EDM ($\hat a_A$). Furthermore,
the EDM is particularly interesting because it is very sensitive to possible new sources of CP violation in the quark and lepton sectors.

In this paper, we have studied the potential of the FCC-he for sensitivity measuring expected on the electromagnetic anomalous couplings
in the $t\bar t \gamma$ vertex. We consider the single top-quark production mode $e^-p \to e^- \bar b \to \bar t \nu_e \gamma$ with $W$
boson exchange via the $t$-channel, that is through charged current production. This channel has the largest cross-section and is hence
the dominant production mode of single top-quark \cite{Sun}. Our study is based on the projections for the
center-of-mass energies $\sqrt{s}$, the integrated luminosity ${\cal L}$ and the polarization electron beam of the FCC-he.
Additionally, we take into account kinematic cuts and systematic uncertainties $\delta_{sys}=0\%, 3\%, 5\%$. The cut based
optimization at $7.07\hspace{0.8mm}TeV$ and $10\hspace{0.8mm}TeV$, involves a set of selection cuts in various
kinematic variables, carefully chosen with the criterion of not being built with kinematic properties of one or part of the
decay products of the top-quark, that could in principle bias the sensitivity to the anomalous couplings. These kinematic variables
selected for these cuts include $\eta^{\gamma}$, $p^\gamma_T$ and $p^{(\nu)}_T$ as defined in Eq. (12). The final results
of this optimization for the resulting cross-section of the $e^-p \to e^- \bar b \to \bar t \nu_e \gamma$ signal and
prospective sensitivities for the dipole moments $\hat a_V$ and $\hat a_A$ indicated in Figs. 3-10 and Tables II-VII
as well as in Eqs. (14)-(25) and (27)-(38) imply that the process $e^-p  \to \bar t \nu_e\gamma$ at FCC-he is an excellent
option for probing the physics of the top-quark. This makes a future $e^-p$ collider an ideal tool to study the electromagnetic
properties of the top-quark through the $t\bar t \gamma$ vertex.

From Tables III, V and VII, the sensitivity estimated on dipole moments of the top-quark are
$\hat a_V=[-0.2308, 0.2204]$, $|\hat a_A|=0.2259$ at $95\%$ C.L. in the hadronic channel with unpolarized electron beam $P_{e^-}=0\%$.
In the case with polarized electron beam for $P_{e^-}=80\%$ and $P_{e^-}=-80\%$ are $\hat a_V=[-0.3428, 0.3321]$, $|\hat a_A|=0.3371$
and $\hat a_V=[-0.2041, 0.1858]$, $|\hat a_A|=0.1939$ at $95\%$ C.L. The corresponding results for the leptonic channel with
$P_{e^-}=0\%, 80\% -80\%$ are $\hat a_V=[-0.3067, 0.2963]$, $|\hat a_A|=0.3019$, $\hat a_V=[-0.4563, 0.4456]$, $|\hat a_A|=0.4505$
and $\hat a_V=[-0.2695, 0.2512]$, $|\hat a_A|=0.2592$, respectively. The results for $\hat a_V$ and $\hat a_A$ in the leptonic channel
are weaker by a factor of 0.75 than those corresponding to the hadronic channel. From these results, we find that the sensitivity estimated
on dipole moments of the top-quark are of the same order of magnitude as those reported in Table I and Refs. \cite{Ibrahim,Atwood,Polouse,Choi,Polouse1,Aguilar0,Amjad,Juste,Asner,Abe,Aarons,Brau,Baer,Grzadkowski:2005ye,murat,Billur}.
In particular, of the comparison with the constraints obtained from the radiative $b \to s\gamma$ transitions at Tevatron and LHC \cite{Bouzas}
and the process $pp\to p\gamma^* \gamma^* p \to pt\bar t p$ at LHC \cite{Sh}, our results are more sensitive. Given these prospective
sensitivities, we highlight that the FCC-he is the potential top-quark factory that is particularly well suited to sensitivity study
on its dipole moments with cleaner environments.

Summarizing, the FCC-he offers us significant opportunities to study the anomalous couplings of the quark-top. However, more
extensive studies on the theoretical, phenomenological and experimental level they are needed. These new possibilities for
investigating the electromagnetic properties of the top-quark will eventually open new avenues in the understanding of the
quark-top physics, as well as new physics BSM.

\vspace{1.5cm}

\begin{center}
{\bf Acknowledgments}
\end{center}

A. G. R. and M. A. H. R. acknowledge support from SNI and PROFOCIE (M\'exico).

\vspace{1.5cm}

\begin{table}[!ht]
\caption{Sensitivities on the AMDM $\hat a_V$ and the EDM $\hat a_A$ of the top-quark through the process
$e^-p  \to e^-\bar b \to \bar t \nu_e\gamma$ at the FCC-he.}
\begin{center}
 \begin{tabular}{|cc|cc|cc|}
\hline\hline
\multicolumn{6}{|c|}{ $\sqrt{s}=$ 7.07 TeV,  \hspace{5mm}  $P_{e^-} = 0 \%$, \hspace{5mm}  $95\%$ C.L.} \\
\hline
\multicolumn{2}{|c|}{} & \multicolumn{2}{c|}{Hadronic} & \multicolumn{2}{c|}{Leptonic} \\
\hline
\cline{1-6}
${\cal L} \, (fb^{-1})$  & \hspace{0.5cm} $ \delta_{sys}$ \hspace{0.5cm}  &
\hspace{1.5cm} $\hat a_V$ \hspace{1.5cm} &
\hspace{0.5cm} $|\hat a_A|$ \hspace{0.5cm}  &
\hspace{1.5cm} $\hat a_V$ \hspace{1.5cm}  &
\hspace{0.5cm} $|\hat a_A|$ \hspace{0.5cm} \\
\hline
50  &  $0\%$   & [-0.6599, 0.6578]    &    0.6587  &  [-0.8815, 0.8794] &  0.8802  \\
50  &  $3\%$   & [-1.3758, 1.3737]    &    1.3742  &  [-1.4138, 1.4118] &  1.4123  \\
50  &  $5\%$   & [-1.7606, 1.7585]    &    1.7589  &  [-1.7792, 1.7772] &  1.7776  \\
\hline
100 &  $0\%$   & [-0.5551, 0.5530]    &    0.5539  &  [-0.7414, 0.7393] &  0.7401  \\
100 &  $3\%$   & [-1.3666, 1.3645]    &    1.3651  &  [-1.3864, 1.3843] &  1.3849  \\
100 &  $5\%$   & [-1.7563, 1.7542]    &    1.7546  &  [-1.7657, 1.7636] &  1.7640  \\
\hline
300  &  $0\%$   & [-0.4221, 0.4199]   &    0.4208  &  [-0.5636, 0.5615] &  0.5624  \\
300  &  $3\%$   & [-1.3604, 1.3583]   &    1.3589  &  [-1.3672, 1.3651] &  1.3656  \\
300  &  $5\%$   & [-1.7534, 1.7513]   &    1.7517  &  [-1.7565, 1.7545] &  1.7549  \\
\hline
500 &  $0\%$   & [-0.3715, 0.3695]    &    0.3704  &  [-0.4962, 0.4941] &  0.4949  \\
500 &  $3\%$   & [-1.3591, 1.3571]    &    1.3576  &  [-1.3632, 1.3612] &  1.3617  \\
500 &  $5\%$   & [-1.7528, 1.7507]    &    1.7511  &  [-1.7547, 1.7526] &  1.7530  \\
\hline
1000 &  $0\%$   & [-0.3126, 0.3105]   &    0.3115  &  [-0.4174, 0.4153] &  0.4162  \\
1000 &  $3\%$   & [-1.3582, 1.3561]   &    1.3567  &  [-1.3602, 1.3582] &  1.3587  \\
1000 &  $5\%$   & [-1.7523, 1.7503]   &    1.7507  &  [-1.7533, 1.7512] &  1.7516  \\
\hline\hline
\end{tabular}
\end{center}
\end{table}

\begin{table}[!ht]
\caption{Sensitivities on the AMDM $\hat a_V$ and the EDM $\hat a_A$ of the top-quark
through the process $e^-p  \to e^-\bar b \to \bar t \nu_e\gamma$ at the FCC-he.}
\begin{center}
 \begin{tabular}{|cc|cc|cc|}
\hline\hline
\multicolumn{6}{|c|}{ $\sqrt{s}=$ 10 TeV, \hspace{5mm} $P_{e^-} = 0 \%$, \hspace{5mm}  $95\%$ C.L.} \\
\hline
\multicolumn{2}{|c|}{} & \multicolumn{2}{c|}{Hadronic} & \multicolumn{2}{c|}{Leptonic} \\
\hline
\cline{1-6}
${\cal L} \, (fb^{-1})$  & \hspace{0.5cm} $ \delta_{sys}$ \hspace{0.5cm}  &
\hspace{1.5cm} $\hat a_V$ \hspace{1.5cm} &
\hspace{0.5cm} $|\hat a_A|$ \hspace{0.5cm}  &
\hspace{1.5cm} $\hat a_V$ \hspace{1.5cm}  &
\hspace{0.5cm} $|\hat a_A|$ \hspace{0.5cm} \\
\hline
50  &  $0\%$   & [-0.4824, 0.4719] & 0.4777 & [-0.6428, 0.6324] & 0.6384 \\
50  &  $3\%$   & [-1.1429, 1.1326] & 1.1391 & [-1.1618, 1.1514] & 1.1579 \\
50  &  $5\%$   & [-1.4667, 1.4563] & 1.4632 & [-1.4757, 1.4653] & 1.4722 \\
\hline
100 &  $0\%$   & [-0.4065, 0.3961] & 0.4017 & [-0.5414, 0.5309] & 0.5368 \\
100 &  $3\%$   & [-1.1386, 1.1281] & 1.1347 & [-1.1482, 1.1378] & 1.1443 \\
100 &  $5\%$   & [-1.4647, 1.4542] & 1.4612 & [-1.4692, 1.4588] & 1.4657 \\
\hline
300 &  $0\%$   & [-0.3101, 0.2997] & 0.3052 & [-0.4126, 0.4022] & 0.4079 \\
300 &  $3\%$   & [-1.1356, 1.1252] & 1.1317 & [-1.1388, 1.1284] & 1.1349 \\
300 &  $5\%$   & [-1.4633, 1.4529] & 1.4598 & [-1.4648, 1.4544] & 1.4613 \\
\hline
500 &  $0\%$   & [-0.2735, 0.2631] & 0.2686 & [-0.3638, 0.3534] & 0.3589 \\
500 &  $3\%$   & [-1.1349, 1.1246] & 1.1311 & [-1.1369, 1.1265] & 1.1330 \\
500 &  $5\%$   & [-1.4629, 1.4526] & 1.4595 & [-1.4639, 1.4535] & 1.4604 \\
\hline
1000 &  $0\%$   & [-0.2308, 0.2204] & 0.2259 & [-0.3067, 0.2963] & 0.3019 \\
1000 &  $3\%$   & [-1.1345, 1.1241] & 1.1306 & [-1.1355, 1.1251] & 1.1316 \\
1000 &  $5\%$   & [-1.4628, 1.4524] & 1.4593 & [-1.4632, 1.4528] & 1.4597 \\
\hline\hline
\end{tabular}
\end{center}
\end{table}


\begin{table}
\caption{Sensitivities on the AMDM $\hat a_V$ and the EDM $\hat a_A$ of the top-quark
through the process $e^-p  \to e^-\bar b \to \bar t \nu_e\gamma$ at the FCC-he.}
\begin{center}
\begin{tabular}{|cc|cc|cc|}
\hline\hline
\multicolumn{6}{|c|}{$\sqrt{s}=$ 7.07 TeV, \hspace{5mm} $P_e = -80 \%$, \hspace{5mm}  $95\%$ C.L.} \\
\hline
\multicolumn{2}{|c|}{} & \multicolumn{2}{c|}{Hadronic} & \multicolumn{2}{c|}{Leptonic} \\
\hline
\cline{1-6}
${\cal L} \, (fb^{-1})$  & \hspace{0.5cm} $ \delta_{sys}$ \hspace{0.5cm}  &
\hspace{1.5cm} $\hat a_V$ \hspace{1.5cm} &
\hspace{0.5cm} $|\hat a_A|$ \hspace{0.5cm}  &
\hspace{1.5cm} $\hat a_V$ \hspace{1.5cm}  &
\hspace{0.5cm} $|\hat a_A|$ \hspace{0.5cm} \\
\hline
50  &  $0\%$   & [-0.5756, 0.5574]   &    0.5696  & [-0.7694, 0.7512]  & 0.7612  \\
50  &  $3\%$   & [-1.3809, 1.3627]   &    1.3685  & [-1.4029, 1.3848]  & 1.3904  \\
50  &  $5\%$   & [-1.7727, 1.7545]   &    1.7582  & [-1.7832, 1.7651]  & 1.7687  \\
\hline
100 &  $0\%$   & [-0.4835, 0.4654]   &    0.4789  & [-0.6469, 0.6288]  & 0.6401  \\
100 &  $3\%$   & [-1.3757, 1.3575]   &    1.3633  & [-1.3869, 1.3688]  & 1.3746  \\
100 &  $5\%$   & [-1.7702, 1.7521]   &    1.7558  & [-1.7756, 1.7574]  & 1.7611  \\
\hline
300 &  $0\%$   & [-0.3659, 0.3477]   &    0.3639  & [-0.4910, 0.4729]  & 0.4864  \\
300 &  $3\%$   & [-1.3722, 1.3540]   &    1.3599  & [-1.3760, 1.3579]  & 1.3637  \\
300 &  $5\%$   & [-1.7686, 1.7504]   &    1.7541  & [-1.7704, 1.7522]  & 1.7559  \\
\hline
500 &  $0\%$   & [-0.3209, 0.3027]   &    0.3203  & [-0.4316, 0.4134]  & 0.4280  \\
500 &  $3\%$   & [-1.3715, 1.3533]   &    1.3592  & [-1.3738, 1.3556]  & 1.3615  \\
500 &  $5\%$   & [-1.7683, 1.7501]   &    1.7538  & [-1.7693, 1.7512]  & 1.7549  \\
\hline
1000 &  $0\%$  & [-0.2678, 0.2496]   &    0.2694  & [-0.3618, 0.3436]  & 0.3599  \\
1000 &  $3\%$  & [-1.3709, 1.3528]   &    1.3586  & [-1.3721, 1.3539]  & 1.3598  \\
1000 &  $5\%$  & [-1.7680, 1.7499]   &    1.7536  & [-1.7686, 1.7504]  & 1.7541  \\
\hline\hline
\end{tabular}
\end{center}
\end{table}

\begin{table}
\caption{Sensitivities on the AMDM $\hat a_V$ and the EDM $\hat a_A$ of the top-quark
through the process $e^-p  \to e^-\bar b \to \bar t \nu_e\gamma$ at the FCC-he.}
\begin{center}
\begin{tabular}{|cc|cc|cc|}
\hline\hline
\multicolumn{6}{|c|}{ $\sqrt{s}=$ 10 TeV, \hspace{5mm} $P_e = -80 \%$, \hspace{5mm}  $95\%$ C.L.} \\
\hline
\multicolumn{2}{|c|}{} & \multicolumn{2}{c|}{Hadronic} & \multicolumn{2}{c|}{Leptonic} \\
\hline
\cline{1-6}
${\cal L} \, (fb^{-1})$  & \hspace{0.5cm} $ \delta_{sys}$ \hspace{0.5cm}  &
\hspace{1.5cm} $\hat a_V$ \hspace{1.5cm} &
\hspace{0.5cm} $|\hat a_A|$ \hspace{0.5cm}  &
\hspace{1.5cm} $\hat a_V$ \hspace{1.5cm}  &
\hspace{0.5cm} $|\hat a_A|$ \hspace{0.5cm} \\
\hline
50  &  $0\%$   & [-0.4210, 0.4027]   &    0.4103  & [-0.5595, 0.5411]  & 0.5482  \\
50  &  $3\%$   & [-1.1423, 1.1239]   &    1.1287  & [-1.1529, 1.1346]  & 1.1394  \\
50  &  $5\%$   & [-1.4679, 1.4496]   &    1.4531  & [-1.4729, 1.4546]  & 1.4581  \\
\hline
100 &  $0\%$   & [-0.3556, 0.3372]   &    0.3450  & [-0.4719, 0.4536]  & 0.4609  \\
100 &  $3\%$   & [-1.1399, 1.1215]   &    1.1263  & [-1.1453, 1.1269]  & 1.1317  \\
100 &  $5\%$   & [-1.4668, 1.4484]   &    1.4520  & [-1.4693, 1.4509]  & 1.4545  \\
\hline
300 &  $0\%$   & [-0.2724, 0.2541]   &    0.2621  & [-0.3609, 0.3425]  & 0.3503  \\
300 &  $3\%$   & [-1.1382, 1.1198]   &    1.1246  & [-1.1400, 1.1216]  & 1.1264  \\
300 &  $5\%$   & [-1.4660, 1.4477]   &    1.4512  & [-1.4669, 1.4485]  & 1.4520  \\
\hline
500 &  $0\%$   & [-0.2409, 0.2226]   &    0.2307  & [-0.3187, 0.3004]  & 0.3083  \\
500 &  $3\%$   & [-1.1379, 1.1195]   &    1.1243  & [-1.1389, 1.1206]  & 1.1254  \\
500 &  $5\%$   & [-1.4659, 1.4475]   &    1.4510  & [-1.4664, 1.4480]  & 1.4515  \\
\hline
1000 &  $0\%$  & [-0.2041, 0.1858]   &    0.1939  & [-0.2695, 0.2512]  & 0.2592  \\
1000 &  $3\%$  & [-1.1376, 1.1192]   &    1.1240  & [-1.1382, 1.1198]  & 1.1246  \\
1000 &  $5\%$  & [-1.4657, 1.4474]   &    1.4509  & [-1.4660, 1.4476]  & 1.4512  \\
\hline\hline
\end{tabular}
\end{center}
\end{table}

\begin{table}
\caption{Sensitivities on the AMDM $\hat a_V$ and the EDM $\hat a_A$ of the top-quark
through the process $e^-p  \to e^-\bar b \to \bar t \nu_e\gamma$ at the FCC-he.}
\begin{center}
\begin{tabular}{|cc|cc|cc|}
\hline\hline
\multicolumn{6}{|c|}{ $\sqrt{s}=$ 7.07 TeV, \hspace{5mm} $P_e = 80 \%$, \hspace{5mm}  $95\%$ C.L.} \\
\hline
\multicolumn{2}{|c|}{} & \multicolumn{2}{c|}{Hadronic} & \multicolumn{2}{c|}{Leptonic} \\
\hline
\cline{1-6}
${\cal L} \, (fb^{-1})$  & \hspace{0.5cm} $ \delta_{sys}$ \hspace{0.5cm}  &
\hspace{1.5cm} $\hat a_V$ \hspace{1.5cm} &
\hspace{0.5cm} $|\hat a_A|$ \hspace{0.5cm}  &
\hspace{1.5cm} $\hat a_V$ \hspace{1.5cm}  &
\hspace{0.5cm} $|\hat a_A|$ \hspace{0.5cm} \\
\hline
50  &  $0\%$   & [-0.9861, 0.9838]   &    0.9842  & [-1.3174, 1.3150]  & 1.3152  \\
50  &  $3\%$   & [-1.4429, 1.4406]   &    1.4406  & [-1.5905, 1.5882]  & 1.5881  \\
50  &  $5\%$   & [-1.7936, 1.7915]   &    1.7913  & [-1.8772, 1.8749]  & 1.8746  \\
\hline
100 &  $0\%$   & [-0.8294, 0.8271]   &    0.8276  & [-1.1079, 1.1056]  & 1.1059  \\
100 &  $3\%$   & [-1.4019, 1.3996]   &    1.3997  & [-1.4874, 1.4851]  & 1.4852  \\
100 &  $5\%$   & [-1.7731, 1.7708]   &    1.7705  & [-1.8177, 1.8153]  & 1.8151  \\
\hline
300 &  $0\%$   & [-0.6305, 0.6282]   &    0.6289  & [-0.8421, 0.8398]  & 0.8404  \\
300 &  $3\%$   & [-1.3725, 1.3702]   &    1.3702  & [-1.4046, 1.4023]  & 1.4024  \\
300 &  $5\%$   & [-1.7588, 1.7565]   &    1.7563  & [-1.7744, 1.7721]  & 1.7719  \\
\hline
500 &  $0\%$   & [-0.5550, 0.5527]   &    0.5535  & [-0.7413, 0.7389]  & 0.7396  \\
500 &  $3\%$   & [-1.3663, 1.3640]   &    1.3641  & [-1.3861, 1.3838]  & 1.3839  \\
500 &  $5\%$   & [-1.7559, 1.7536]   &    1.7534  & [-1.7654, 1.7630]  & 1.7628  \\
\hline
1000 &  $0\%$  & [-0.4669, 0.4646]   &    0.4654  & [-0.6236, 0.6212]  & 0.6219  \\
1000 &  $3\%$  & [-1.3617, 1.3594]   &    1.3595  & [-1.3718, 1.3695]  & 1.3696  \\
1000 &  $5\%$  & [-1.7537, 1.7514]   &    1.7512  & [-1.7585, 1.7562]  & 1.7559  \\
\hline\hline
\end{tabular}
\end{center}
\end{table}

\begin{table}
\caption{Sensitivities on the AMDM $\hat a_V$ and the EDM $\hat a_A$ of the top-quark
through the process $e^-p  \to e^-\bar b \to \bar t \nu_e\gamma$ at the FCC-he.}
\begin{center}
\begin{tabular}{|cc|cc|cc|}
\hline\hline
\multicolumn{6}{|c|}{ $\sqrt{s}=$ 10 TeV, \hspace{5mm} $P_e = 80 \%$, \hspace{5mm} $95\%$ C.L.} \\
\hline
\multicolumn{2}{|c|}{} & \multicolumn{2}{c|}{Hadronic} & \multicolumn{2}{c|}{Leptonic} \\
\hline
\cline{1-6}
${\cal L} \, (fb^{-1})$  & \hspace{0.5cm} $ \delta_{sys}$ \hspace{0.5cm}  &
\hspace{1.5cm} $\hat a_V$ \hspace{1.5cm} &
\hspace{0.5cm} $|\hat a_A|$ \hspace{0.5cm}  &
\hspace{1.5cm} $\hat a_V$ \hspace{1.5cm}  &
\hspace{0.5cm} $|\hat a_A|$ \hspace{0.5cm} \\
\hline
50  &  $0\%$   & [-0.7189, 0.7083]   &    0.7129  & [-0.9589, 0.9482]  & 0.9526  \\
50  &  $3\%$   & [-1.1770, 1.1663]   &    1.1703  & [-1.2567, 1.2460]  & 1.2499  \\
50  &  $5\%$   & [-1.4835, 1.4728]   &    1.4765  & [-1.5256, 1.5149]  & 1.5185  \\
\hline
100 &  $0\%$   & [-0.6054, 0.5947]   &    0.5995  & [-0.8072, 0.7965]  & 0.8011  \\
100 &  $3\%$   & [-1.1563, 1.1456]   &    1.1497  & [-1.2003, 1.1896]  & 1.1936  \\
100 &  $5\%$   & [-1.4733, 1.4627]   &    1.4663  & [-1.4953, 1.4846]  & 1.4882  \\
\hline
300 &  $0\%$   & [-0.4613, 0.4506]   &    0.4555  & [-0.6146, 0.6039]  & 0.6087  \\
300 &  $3\%$   & [-1.1419, 1.1312]   &    1.1352  & [-1.1576, 1.1469]  & 1.1509  \\
300 &  $5\%$   & [-1.4665, 1.4558]   &    1.4594  & [-1.4739, 1.4633]  & 1.4669  \\
\hline
500 &  $0\%$   & [-0.4067, 0.3959]   &    0.4009  & [-0.5416, 0.5308]  & 0.5357  \\
500 &  $3\%$   & [-1.1389, 1.1282]   &    1.1323  & [-1.1485, 1.1378]  & 1.1419  \\
500 &  $5\%$   & [-1.4651, 1.4544]   &    1.4581  & [-1.4696, 1.4589]  & 1.4626  \\
\hline
1000 &  $0\%$  & [-0.3428, 0.3321]   &    0.3371  & [-0.4563, 0.4456]  & 0.4505  \\
1000 &  $3\%$  & [-1.1367, 1.1259]   &    1.1300  & [-1.1415, 1.1309]  & 1.1348  \\
1000 &  $5\%$  & [-1.4640, 1.4533]   &    1.4570  & [-1.4663, 1.4556]  & 1.4593  \\
\hline\hline
\end{tabular}
\end{center}
\end{table}


\pagebreak

\begin{figure}[t]
\centerline{\scalebox{0.8}{\includegraphics{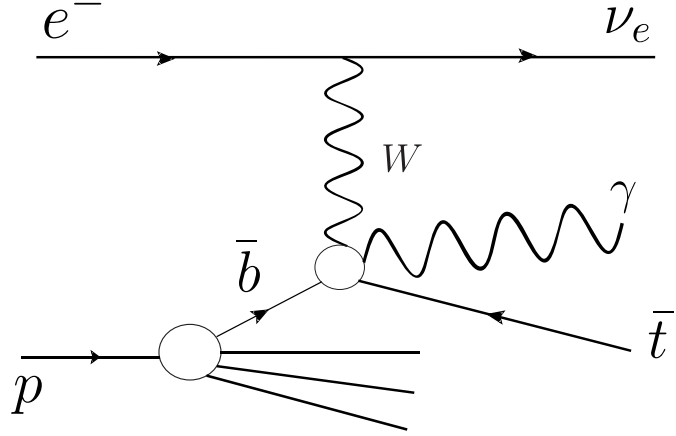}}}
\caption{ \label{fig:gamma1} A schematic diagram for the
single top-quark production through charged currents at
$e^-p$ colliders via the process $e^-p \to e^- \bar b \to \bar t \nu_e \gamma$.}
\label{Fig.1}
\end{figure}

\begin{figure}[t]
\centerline{\scalebox{0.8}{\includegraphics{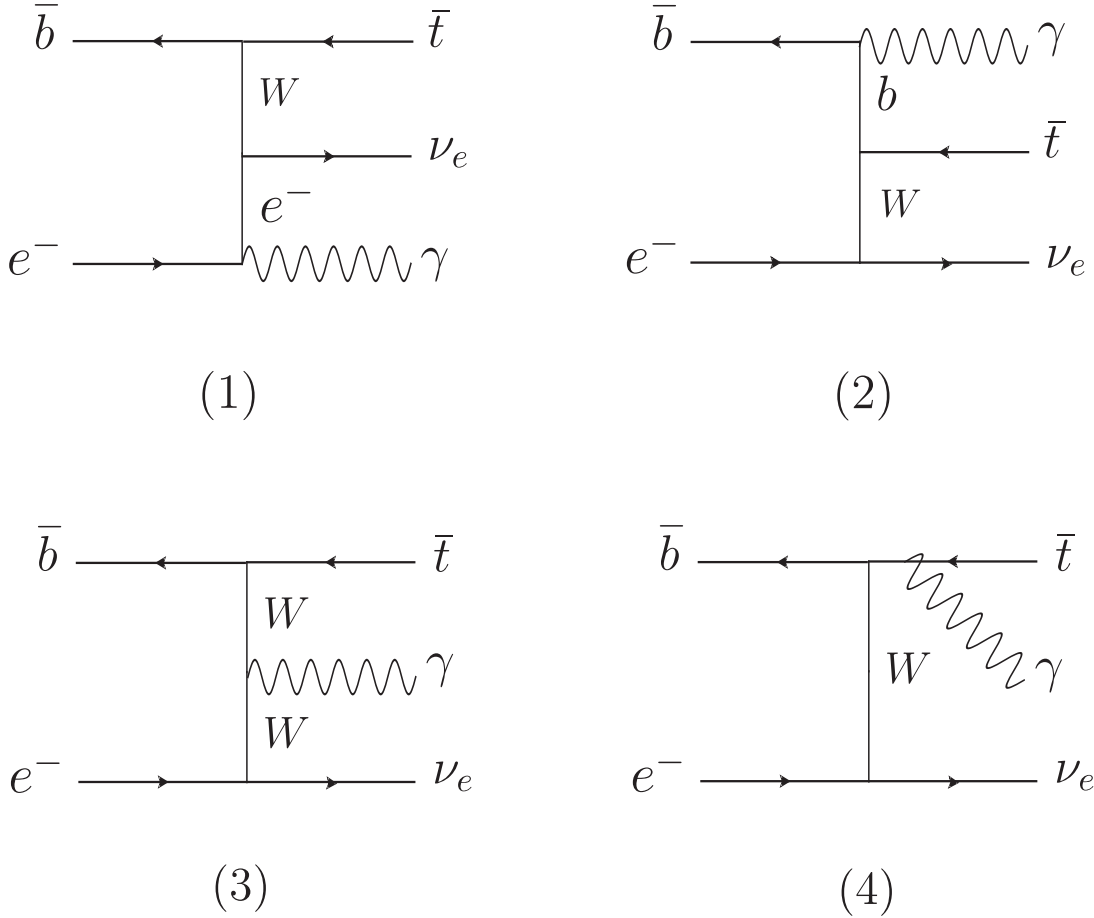}}}
\caption{ \label{fig:gamma2} Feynman diagrams contributing
single top-quark production through charge current at $e^-p$
collider via the subprocess $e^- \bar b \to \bar t \nu_e \gamma $.}
\label{Fig.2}
\end{figure}

\begin{figure}[t]
\centerline{\scalebox{1.2}{\includegraphics{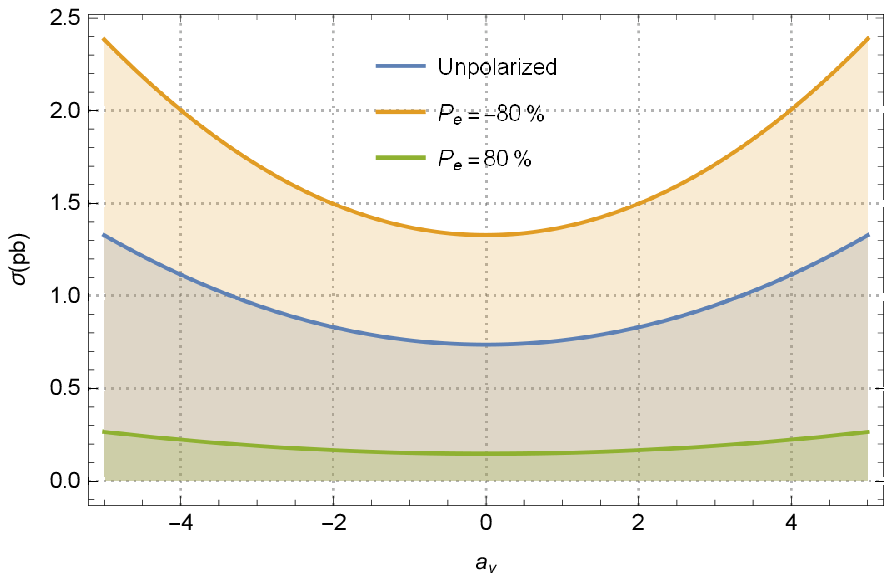}}}
\caption{ The total cross sections of the process
$e^-p  \to e^-\bar b \to \bar t \nu_e\gamma$ as a function of $\hat a_V$
for center-of-mass energies of $\sqrt{s}=7.07$\hspace{0.8mm}$TeV$ at the FCC-he.}
\label{Fig.3}
\end{figure}

\begin{figure}[t]
\centerline{\scalebox{1.2}{\includegraphics{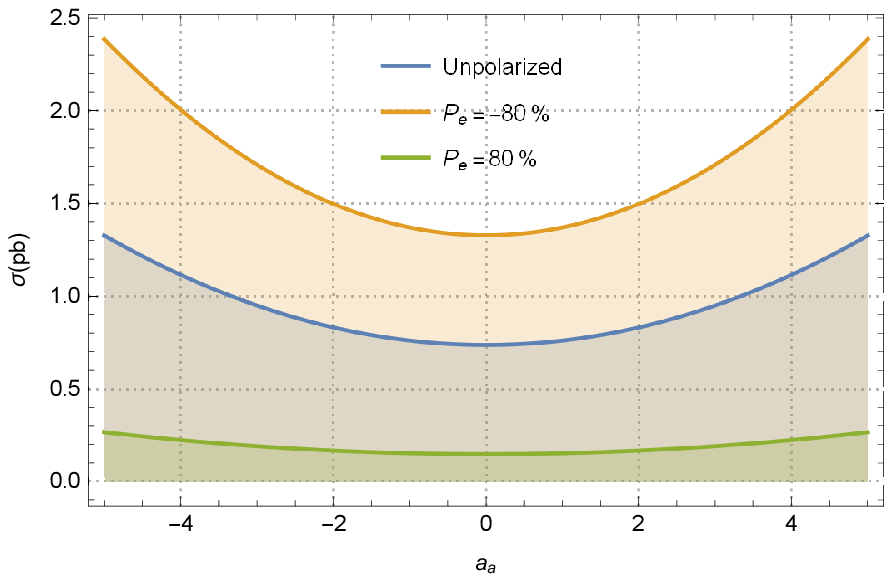}}}
\caption{ The total cross sections of the process
$e^-p  \to e^-\bar b \to \bar t \nu_e\gamma$ as a function of $\hat a_A$
for center-of-mass energies of $\sqrt{s}=7.07$\hspace{0.8mm}$TeV$ at the FCC-he.}
\label{Fig.3}
\end{figure}

\begin{figure}[t]
\centerline{\scalebox{1.2}{\includegraphics{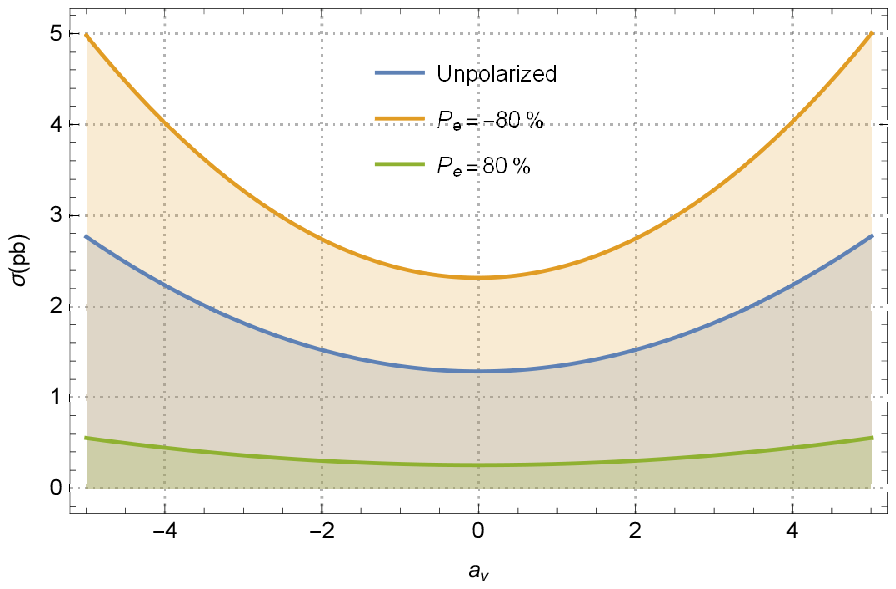}}}
\caption{ Same as in Fig. 3, but for $\sqrt{s}=10$\hspace{0.8mm}$TeV$.}
\label{Fig.4}
\end{figure}

\begin{figure}[t]
\centerline{\scalebox{1.2}{\includegraphics{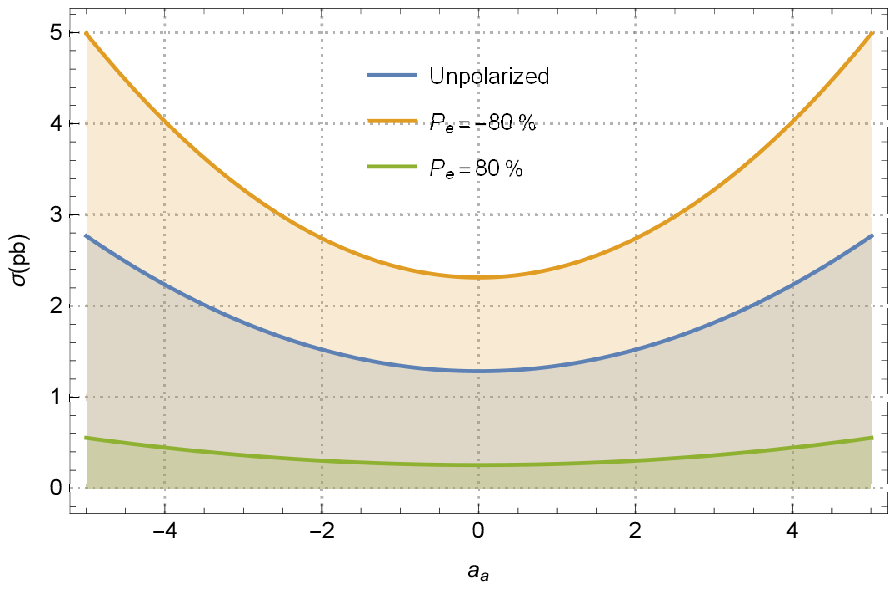}}}
\caption{ Same as in Fig. 4, but for $\sqrt{s}=10$\hspace{0.8mm}$TeV$.}
\label{Fig.4}
\end{figure}

\begin{figure}[t]
\centerline{\scalebox{1.4}{\includegraphics{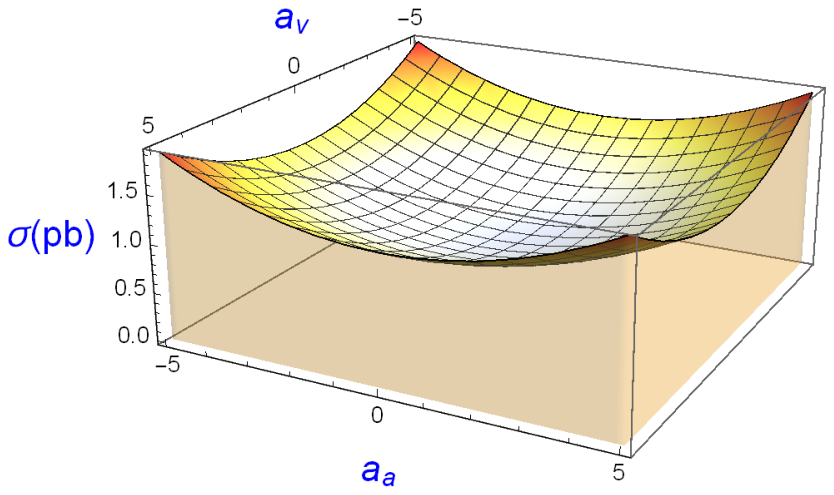}}}
\caption{ \label{fig:gamma1} The total cross sections of the process
$e^-p  \to e^-\bar b \to \bar t \nu_e\gamma$ as a function of $\hat a_V$ and $\hat a_A$
for center-of-mass energy of $\sqrt{s}=7.07$\hspace{0.8mm}$TeV$ at the FCC-he.}
\label{Fig.5}
\end{figure}

\begin{figure}[t]
\centerline{\scalebox{1.4}{\includegraphics{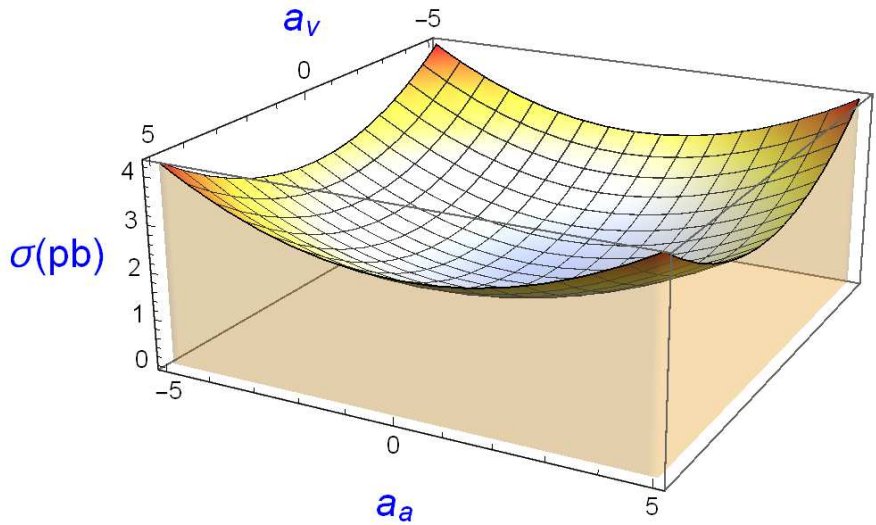}}}
\caption{ \label{fig:gamma2} Same as in Fig. 7, but for center-of-mass energy of
$\sqrt{s}=10$\hspace{0.8mm}$TeV$.}
\label{Fig.6}
\end{figure}

\begin{figure}[t]
\centerline{\scalebox{1.1}{\includegraphics{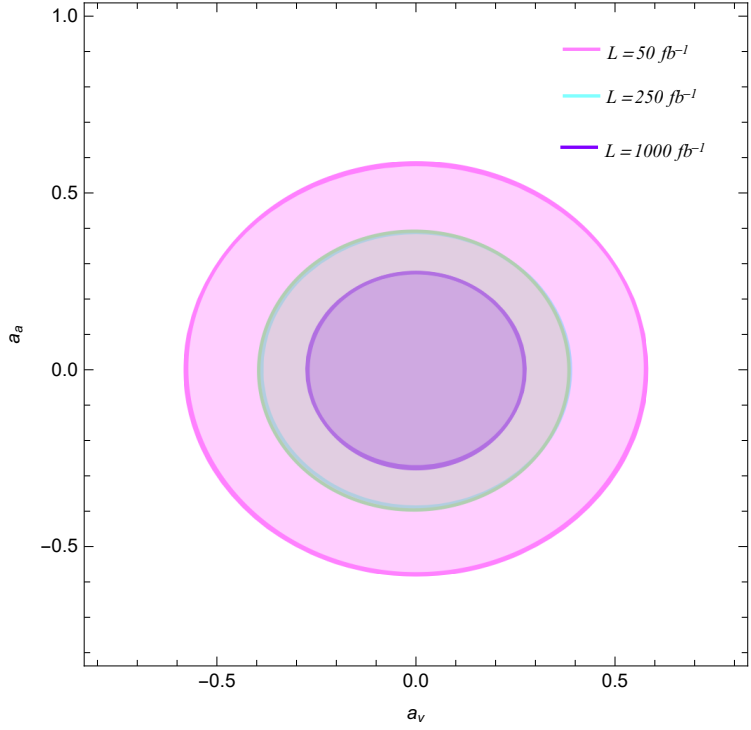}}}
\caption{ \label{fig:gamma1x} Sensitivity contours at the $95\% \hspace{1mm}C.L.$ in the
$\hat a_V-\hat a_A$ plane through the process $e^-p  \to e^-\bar b \to \bar t \nu_e\gamma$
for $\sqrt{s}=7.07$\hspace{0.8mm}$TeV$ at the FCC-he.}
\label{Fig.7}
\end{figure}

\begin{figure}[t]
\centerline{\scalebox{1.1}{\includegraphics{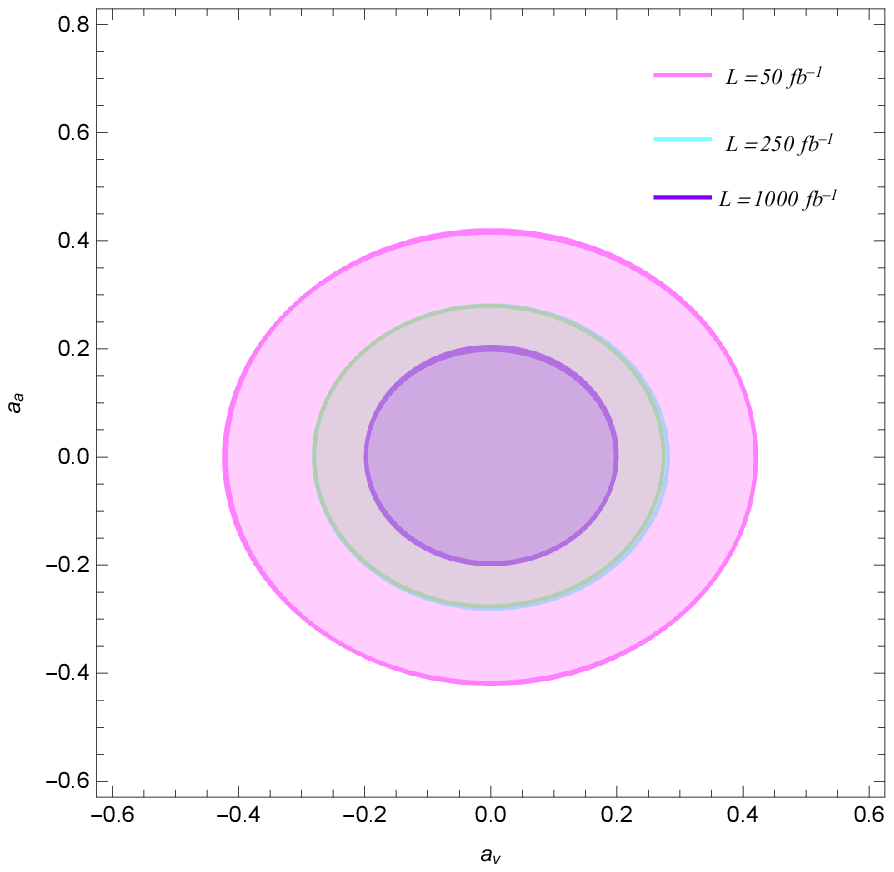}}}
\caption{ \label{fig:gamma2x} Same as in Fig. 9, but for $\sqrt{s}=10$\hspace{0.8mm}$TeV$.}
\label{Fig.8}
\end{figure}

\end{document}